\documentclass{elsart}

\usepackage[latin1]{inputenc}
\usepackage[T1]{fontenc}
\usepackage{times}
\usepackage{amsmath}
\usepackage{amsfonts}
\usepackage{amssymb}
\usepackage{wasysym}

\usepackage{acronym}
\usepackage{comment}

\usepackage{fancyhdr}
\usepackage{graphicx}
\usepackage{url}

\usepackage{natbib}

\begin{document}

\begin{frontmatter}
\title{Parameters Affecting the Resilience of Scale-Free Networks to
  Random Failures}

\author[UNM,SNL]{Hamilton Link}
\author[SNL]{Randall A. La{V}iolette}
\author[UNM]{Jared Saia}
\author[UNM]{Terran Lane}
\address[UNM]{Department of Computer Science \\ University of New Mexico \\ Albuquerque, NM  87131}
\address[SNL]{Sandia National Laboratories \\ Albuquerque, NM  87185}

\maketitle

\begin{abstract}
It is commonly believed that scale-free networks are robust to massive
numbers of random node deletions.  For example, Cohen et al.\ in
\citep{cohen00resilience} study scale-free networks including some
which approximate the measured degree distribution of the Internet.
Their results suggest that if each node in this network failed
independently with probability $0.99$, the remaining network would
continue to have a giant component.  In this paper, we show that a
large and important subclass of scale-free networks are \emph{not}
robust to massive numbers of random node deletions for practical
purposes.  In particular, we study finite scale-free networks which
have minimum node degree of $1$ and a power-law degree distribution
beginning with nodes of degree $1$ (power-law networks).  We show
that, in a power-law network approximating the Internet's reported
distribution, when the probability of deletion of each node is $0.5$
only about $25\%$ of the surviving nodes in the network remain
connected in a giant component, and the giant component does not
persist beyond a critical failure rate of $0.9$.  The new result is
partially due to improved analytical accommodation of the large number
of degree-$0$ nodes that result after node deletions.  Our results
apply to finite power-law networks with a wide range of power-law
exponents, including Internet-like networks.  We give both analytical
and empirical evidence that such networks are not generally robust to
massive random node deletions.
\end{abstract}

\end{frontmatter}

\section{Introduction}

Scale-free networks (SFNs) are massive networks whose node-degree
distribution follows a power law in the tail of the distribution
\citep{aiello00random, albert02statistical}.  Power-law networks
(PLNs) are the class of scale-free networks which have a minimum
degree of $1$ and which follow a power law beginning at degree $1$.
Many real-world networks---such as the Internet, the web graph, and
many social networks---have been observed to be scale-free
\citep{faloutsos99power-law, chang04towards, csanyi04structure,
  reed04brief, saroiu02measurement}.  Because of the prevalence of
these networks, the relationship between a network's degree
distribution and its robustness to random node deletions has been
studied, and the common belief is that scale-free networks are very
robust to this kind of failure.  The original work on this subject
\citep{cohen00resilience} led to the claim that the Internet would
retain a giant component even if $99\%$ of its nodes were removed at
random.

To study this desirable property of scale-free networks, we modeled
the effects of random failures on a graph's degree distribution.  This
revealed that power-law networks are {\em not\/} generally robust to
random node failures.  In the case of the widely-cited Internet
resilience result \citep{cohen00resilience} of a power-law network
with slope parameter $\beta = 2.5$ and minimum degree $1$
\citep{faloutsos99power-law, chang04towards}, high failure rates lead
to the orphaning of a large fraction of the surviving nodes, and to
the complete disintegration of the giant component after $90\%$ of the
network has failed.  This high critical failure rate may appear to
suggest robustness, but as the failure rate increases, the giant
component captures a diminishing fraction of the surviving nodes.  For
example, when $\beta = 2.5$, a PLN's giant component initially
represents $60\%$ of the network but comprises only $25\%$ of the
surviving network by the time half the network has failed.  As $\beta$
increases this decay becomes more dramatic, and the critical failure
rate decreases.

The main results of this paper are as follows.  We estimate the
surviving degree distribution of a PLN after random node deletions in
order to capture our simulated results.  The graph that remains after
random node deletions is approximately a PLN, and its degree
distribution can be conservatively estimated with similar parameters.
We show analytically how to derive these parameters from the initial
PLN slope $\beta$ and the failure rate $p$, and use these parameters
to identify the critical failure rate for a PLN.  Our empirical
results validate and expand these analytical results by showing when
simulated PLNs break down and how the giant component decays as a
function of $p$.

We observe that a large and important class of scale-free networks
decays rapidly and has critical failure rates due to finite size
effects, and conclude they are not resilient to massive random
failures.  In practice, dynamic failures are likely to be of more
interest when considering a real network's resilience, but the
simultaneous random failure model is also useful for studying the
structure of subpopulations in a network.  Our result is applicable to
the study of distributed collaborative filters \citep{link05impact},
robust networks, and epidemiology.  If real-world PLNs such as the
Internet and disease pathway networks exhibit robustness, we do not
believe it is because of their power-law distribution, and further
explanations must be sought.  More highly assortative networks with
similar distributions but in which connectivity is biased in favor of
connecting similar nodes \citep{newman03mixing} are worth further
study.

\section{Related Work}\label{section-relatedWork}

A large body of work has been published on graphs and their
properties, and our work builds upon the work in scale-free networks.
The formal treatments are based in physics, statistical mechanics,
computer science, and mathematics \citep{cohen00resilience,
  aiello00random, albert02statistical, molloy95critical, molloy98size,
  otter48number, otter49multiplicative, mohar91laplacian} and describe
mathematical properties of graphs, including those derived from the
assumption that SFN node-degree distributions follow a power law in
the tail. The empirical work \citep{faloutsos99power-law,
  chang04towards, csanyi04structure, reed04brief, saroiu02measurement}
is aimed at capturing or sampling the degree distribution and other
properties of real-world networks such as Internet routing, web pages,
or social networks in order to determine whether the observed systems
are scale free, and to compare observations to the theoretical
properties.  Many authors observe that Internet communities tend to
form scale-free networks, although for the Internet itself this
empirical work is based on traceroute sampling, which has been called
into question \citep{achlioptas05bias}.

Scale-free networks were originally of interest to us because of their
published resilience to random failures \citep{cohen00resilience,
  saroiu02measurement}, which implied that random subpopulations in a
SFN had a good chance of being highly connected.  Our interest in
subpopulations lies in distributed multi-agent systems and distributed
recommender systems, wherein it is desirable that disinterested
parties not be required to process or forward messages
\citep{link05impact, awerbuch05improved}.  The spread of information
and pathogens has also been studied, as have many other families of
graphs \citep{albert02statistical, kempe03maximizing,
  watts98collective, dorogovtsev04shortest}.

It is important to be precise in comparing our work with the result in
\citep{cohen00resilience}.  As part of extremal graph theory,
scale-free and power-law network percolation thresholds are defined if
they hold in the limit as the size of the network goes to infinity.
That result holds in the limit for SFNs with small minimum degree, but
does not hold for finite PLNs of minimum degree $1$ (as many workers
in the field have come to assume).  We seek to correct the
generalization, and here we analyze the effects of finite system size
on the percolation threshold for this special case of SFN.

\section{Random Failures in Power-Law Networks}\label{section-SFNs}

We consider the class of scale-free networks with minimum degree $1$
whose degree distributions begin following a power law at degree $1$.
We refer to these as power-law networks, or PLNs.  PLNs have been
analyzed in some detail \citep{aiello00random, aiello01random} but we
further wish to understand the properties of the subgraph which
remains after random node deletions.  Let the number of nodes of
degree $k$ in an initial graph $G$ be $c(G,k) = e^\alpha k^{-\beta}$,
with power-law slope $\beta$ ($0 < \beta < \beta_0 = 3.47875...$),
scale parameter $\alpha$, and minimum degree $1$
\citep{aiello00random}.  For PLNs there is no giant component when
$\beta > \beta_0$ \citep{aiello00random}.  We denote the total number
of vertices of degree $k \ge 1$ in $G$ as $|G|$, and count degree-$0$
nodes separately as they appear.  Given the parameters for $G$, and a
failure probability $p$, it is reasonable to ask whether the surviving
graph $G'$ is a PLN.  If so, we wish to know its corresponding
parameters $\alpha '$ and $\beta '$, and whether $G'$ has a giant
component (i.e., a connected portion of the graph with $\Theta(|G'|)$
nodes).  If $G'$ is a PLN (apart from orphaned nodes) with $\beta' <
\beta_0$ then a giant component almost surely exists, so it suffices
to show when this condition holds.  Our derivations indicate critical
failure rates ($p$ such that $\beta' = \beta_0$) only appear when
$\beta > 2$, and we restrict ourselves to that case here.

\begin{figure}
\centerline{\mbox{\includegraphics[width=3.5in]{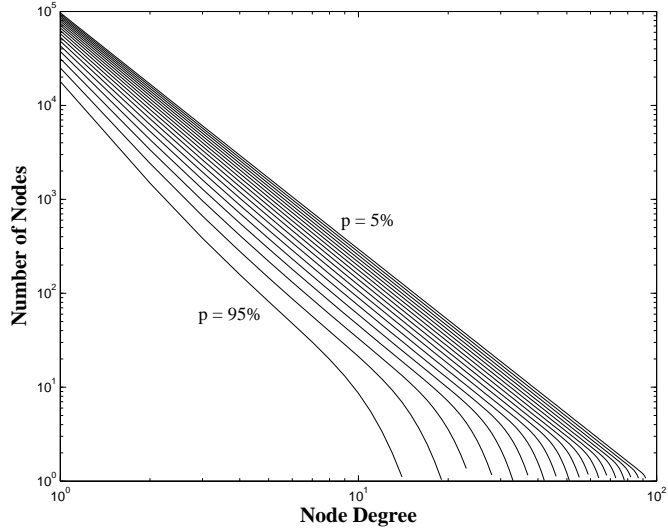}}}
\caption{The degree distribution of the remaining graph when a
  fraction $p = 5\%$--$95\%$ of the nodes in a PLN have failed, for
  $\beta = 2.5$ and $\alpha$ chosen such that the maximum degree was
  $130$.  The graph follows a rough power law with an increase in
  slope and a growing rolloff.}
\label{plotOfCohenResult}
\end{figure}

Our initial numerical and experimental work indicated that when $G$ is
a PLN, the number of surviving nodes in $G'$ of degree $k \ge 1$
follow an approximate power-law degree distribution
(Figure~\ref{plotOfCohenResult}).  In finite networks, these surviving
distributions will exhibit a rolloff.  The observed rolloffs in these
distributions make the graph less robust---the rolloff is comparable
to targeted attacks, which cause SFNs to collapse more quickly
\citep{cohen01breakdown, dorogovtsev01comment}.  Also, the low-degree
behavior of the surviving graph remains very consistent, suggesting
the critical slope will not be affected (low-degree behavior such as
raising the minimum degree etc. can dramatically increase $\beta_0$ or
eliminate the critical slope entirely).  Using a power law
approximation with no rolloff and comparing the slope to the same
critical $\beta_0$ should thus give us a reasonable bound on
robustness to random failures.

The number of nodes $|G|$ in a PLN with $\beta > 1$ is approximately
$\zeta(\beta) e^\alpha$ \citep{aiello00random}, using the Riemann Zeta
function $\zeta(t) = \Sigma_{n=1}^\infty \frac{1}{n^t}$.  This gives
the expected number of nodes of degree $k \ge 1$ in $G$ and $G'$.  For
$G'$ we will also account for orphaned nodes (nodes which have not
failed but whose neighbors have all failed, leaving them with degree
$0$).  For us this is crucial---orphans should not be considered faulty,
as they are only isolated members of the subpopulation of interest.

Assuming $G'$ is suitably approximated with a
power-law, it remains to derive the parameters of $G'$ and determine
from $\beta$ and $p$ when $\beta ' < \beta_0$.  If failures occur with
failure probability $p$ in a graph with degree distribution $c(G,k) =
e^\alpha k^{-\beta}$ the new degree distribution (tightly bounded
around its expectation) will be
$$c(G',k) = (1-p) \sum_{k_0 = k}^{e^{\frac{\alpha}{\beta}}} \frac{e^\alpha}{k_0^\beta} {k_0 \choose k} (1-p)^k p^{k_0 - k}\mbox{.}$$
Which for degree $0$ and $1$ reduce to
\begin{eqnarray}
c(G',0) = (1-p) e^\alpha \chi\mbox{,} \quad & \chi \stackrel{\mbox{\tiny def}}{=} \sum_{k_0 = 1}^{e^{\frac{\alpha}{\beta}}} \frac{1}{k_0^\beta} p^{k_0}\mbox{, and} \nonumber \\
c(G',1) = (1-p) e^\alpha \xi\mbox{,} \quad & \xi \stackrel{\mbox{\tiny def}}{=} \sum_{k_0 = 1}^{e^{\frac{\alpha}{\beta}}} \frac{1}{k_0^\beta} k_0 (1-p) p^{k_0 - 1}\nonumber
\end{eqnarray}
(noting the definitions of $\chi$ and $\xi$).  Estimating the new
distribution with a power law $c(G',k) \approx e^{\alpha'}k^{-\beta'}$
gives $c(G',1) \approx e^{\alpha'} = (1-p) e^\alpha \xi$ and we
directly obtain $\alpha ' = \alpha + \ln ((1-p) \xi)$.  To find
$\beta'$, note that of the original $|G|$ nodes there are an expected
$p|G|$ failed nodes, $c(G',0)$ orphaned nodes, and $|G'|$ nodes
captured by the size estimate of the new graph given the new
parameters.  For $\beta > 1$,
\begin{eqnarray}
|G| & = & p|G| + |G'| + c(G',0) \nonumber \\
\zeta(\beta)e^\alpha & = & p\zeta(\beta)e^\alpha + \zeta(\beta')e^{\alpha'} + (1-p)e^\alpha \chi \nonumber \\
\zeta(\beta)e^\alpha & = & p\zeta(\beta)e^\alpha + \zeta(\beta')(1-p)e^\alpha \xi + (1-p)e^\alpha \chi \nonumber
\end{eqnarray}
and solving for $\beta '$ gives
\begin{eqnarray}
(1-p)\zeta(\beta)e^\alpha & = & \zeta(\beta')(1-p)e^\alpha \xi + (1-p)e^\alpha \chi \nonumber \\
\zeta(\beta) & = & \zeta(\beta') \xi + \chi \nonumber \\
\beta' & = & \zeta^{-1}\left( \frac{\zeta(\beta) - \chi}{\xi} \right) \mbox{.} \nonumber
\end{eqnarray}

\begin{figure}
\centerline{\mbox{\includegraphics[width=3.5in]{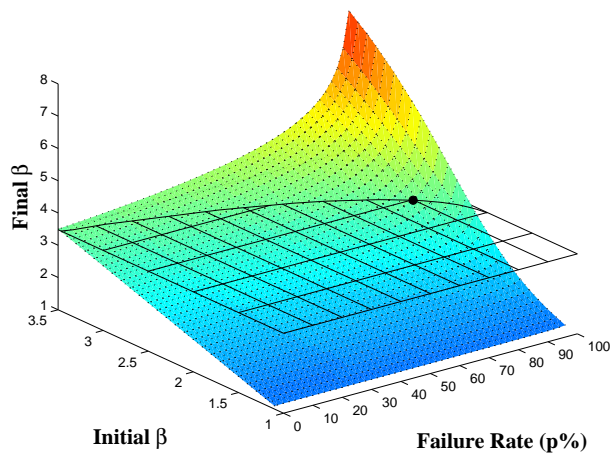}}}
\caption{The function $\beta' (\beta, p)$ for $1.1 \le \beta \le 3.5$ and
  $0 \le p \le 99\% $.  For $\beta' \ge \beta_0$ (intersecting
  plane), no giant component exists in $G'$.  An Internet-like PLN
  with $\beta = 2.5$ will cease to have a giant component when $p =
  89.8\%$ (point emphasized).}
\label{beta-betaPrime}
\end{figure}

\begin{figure}
\centerline{\mbox{\includegraphics[width=3in]{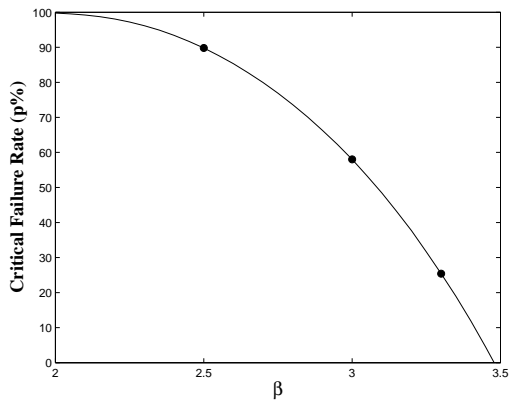}}}
\caption{The critical failure rate $p$ for which $\beta'(\beta,p) =
  \beta_0$, for $2 < \beta < \beta_0$.  The critical points for the
  curves of $\beta = 2.5$ ($p = 89.8\%$), $3.0$ ($p = 58.0\%$), and
  $3.3$ ($p = 25.4\%$) are emphasized for their correspondence with
  empirically studied real networks \citep{csanyi04structure,
    reed04brief, saroiu02measurement}.}
\label{criticalBeta}
\end{figure}

Numerical estimation shows that $\beta' > \beta$, and this difference
varies with $p$.  Figures~\ref{beta-betaPrime} and \ref{criticalBeta}
show that for $\beta > 2$ there are critical failure rates $p_c$ for
which $\beta' = \beta_0$, beyond which the surviving graph will not
have a giant component. This shows that power-law networks are not
generally robust to random node failures.  However, our result depends
upon a potentially coarse approximation and (although we have noted
this approximation should certainly result in an upper bound on $p_c$)
we would like an indication of how accurate our bounds are, and some
validation of our methodology.  The next section will present
additional evidence gathered by observing failures in simulated
graphs.

\section{Simulation Methods and Results}\label{section-sims}

Using $c(G,k) = e^\alpha k^{-\beta}$ we generated node-degree
histograms matching a power law, and recorded ($\beta$, $\alpha$)
pairs that produced histograms averaging one million nodes.  The
histograms were used to populate an array of vertex-numbered ``edge
stubs,'' the stubs were permuted randomly to create a random
configuration \citep{bollobas85random}, and pairs of stubs were added
as edges in a multigraph.

The random configuration produces multigraphs which match a node
degree distribution, but it is reasonable to wonder if redundant arcs
and self-arcs are frequent enough to conflict with the assumptions of
independence implicit in our derivations.  From \citep{aiello00random}
we estimated their likelihood given the number of edges in the graph
and the highest expected degree, and established that they are
infrequent.  For $\beta > 2$ the likelihood of an individual edge of
the highest-degree node being a self-arc is $2 / \zeta(\beta-1)
e^{\frac{\beta-1}{\beta} \alpha}$, which approaches zero in large
PLNs.

We used an iterated 3-coloring algorithm to identify the giant and
secondary components in $G$.  To simulate failures, nodes were
pre-colored with probability $p$ and the algorithm run again to
collect the components of $G'$.  For each $\beta$ and $p$ twenty
independent graphs were created, with mean and standard deviations
collected for several statistics.

\begin{figure}
\centerline{\mbox{\includegraphics[width=3in]{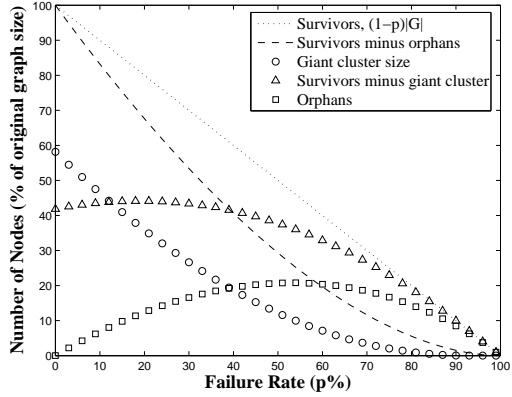}}}
\caption{Simulation results for $\beta = 2.5$, and $\alpha$ chosen
  such that $n \approx 10^6$.  Initial giant component size and
  critical failure rate vary with $\beta$ but this behavior is
  representative of $2 < \beta < \beta_0$.  Note the decreasing
  fractional size of the giant component $\Circle$, and the increasing
  population of orphans $\square$ and nodes in small components in
  general $\vartriangle$.}
\label{internetFailurePlot}
\end{figure}

For $2 \le \beta \le 3.5$ and $0 \le p \le 0.99$ we computed the size
of the first and second largest components, the number of surviving
nodes outside the largest component, and the number of orphans.  The
range was chosen to include $2$, a transition point of interest in
regards to the density of edges in a PLN \citep{aiello00random}, and
$3.5$, to demonstrate random failures in a graph with no giant
component.  For $\beta = 2.5$ the data can be seen in
Figure~\ref{internetFailurePlot}.

\begin{figure}
\centerline{\mbox{\includegraphics[width=3in]{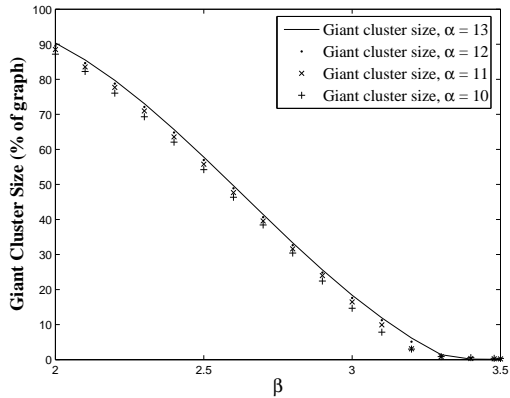}}}
\caption{The fraction of nodes in the giant component (in simulation)
  for $2 \le \beta < \beta_0$, when $n \approx 10^6$.  These numbers
  varied with $\alpha$; for $\beta = 2$ the fraction of nodes in the
  giant component approaches $1$ in the limit \citep{aiello01random},
  but for practical purposes (and possibly in the limit) the giant
  component for $\beta > 2$ can be a small fraction of the graph.}
\label{gcs-fraction}
\end{figure}

For those familiar with the literature, the initial size of the giant
component in Figure~\ref{internetFailurePlot} may come as a surprise.
For $\beta < 2$ the fraction of the graph in the giant component is
indirectly given in \citep{aiello01random} and it is known to comprise
virtually the entire graph.  The probability that a random PLN node is
in the giant component is approximately $1 - \frac{2 \log
  \alpha}{\alpha}$ for $\beta = 2$ and approximately $1 -
\Theta\left(\mbox{exp}[{-\frac{(2-\beta)}
    {(3-\beta)\beta}\alpha}]\right)$ for $\beta < 2$, both of which
approach $1$ in the limit.\footnote{Originally published with a
  typographic error, as confirmed by the authors in
  \citep{aiello05email}.} For $2 < \beta < \beta_0$ no such equation
has been published, but in simulations the fractional size of the
giant component decreases prior to its complete dissolution at
$\beta_0$ as shown in Figure~\ref{gcs-fraction}, subject to some scale
effects.  We have not seen this published elsewhere and the result was
somewhat unexpected, so we include it here.

\begin{figure}
\centerline{\mbox{\includegraphics[width=3.5in]{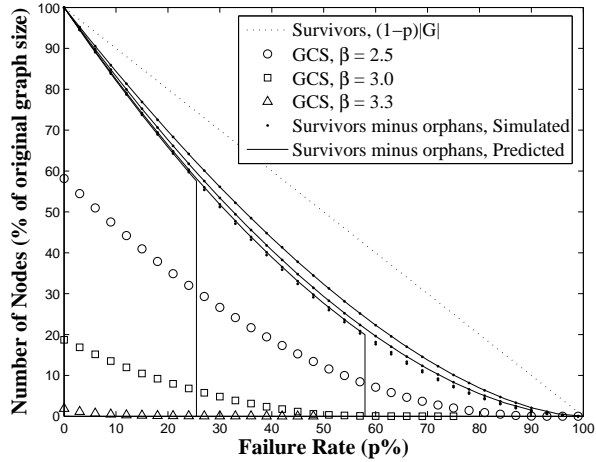}}}
\caption{Predicted and actual unorphaned survivors, predicted critical
  failure rates, and giant component size, for $\beta = 2.0$, $2.5$,
  $3.0$, and $3.3$, as a function of $p$.  The predicted number of
  unorphaned survivors, $(1-p)\zeta(\beta')e^\alpha
  \xi/\zeta(\beta)e^\alpha$, closely matches observed unorphaned
  survivors (as expected of a tightly bounded combinatorial result).
  More importantly, the giant component sizes ($\Circle$ $\square$
  $\vartriangle$) fall away close to the predicted critical failure rates
  (vertical drop).  Giant component sizes for $\beta = 3.0$ and $3.3$
  are truncated for clarity at 75\% and 50\%, respectively.}
\label{sigSizeVsExp}
\end{figure}

Figure~\ref{sigSizeVsExp} presents the central result of our
simulations against our mathematical predictions.  The solid lines
depict $|G'|/|G|$ (i.e.\ $(1-p)\zeta(\beta')e^\alpha
\xi/\zeta(\beta)e^\alpha$ for select values of $\beta$---the
difference between these curves and the ideal (the diagonal,
$(1-p)|G|$) is the number of orphaned nodes.  These curves are
terminated and vertical lines are drawn at the point when $\beta' =
\beta_0$.  Over these curves are plotted pointwise observations of
these quantities in simulations of networks with $10^6$ nodes.  The
match is virtually exact, as one might expect---the combinatorics of
the predictions is simply being exercised stochastically in the
simulation.  Finally and most importantly, Figure~\ref{sigSizeVsExp}
plots the decreasing size of the giant component in the graph for
comparison with the vertically distinguished critical failure rates.
In this case the simulation is being compared to our approximation,
and we see that the giant component falls away to virtually nothing as
the failure rate approaches the predicted critical point.

\begin{figure}
\centerline{\mbox{\includegraphics[width=3.5in]{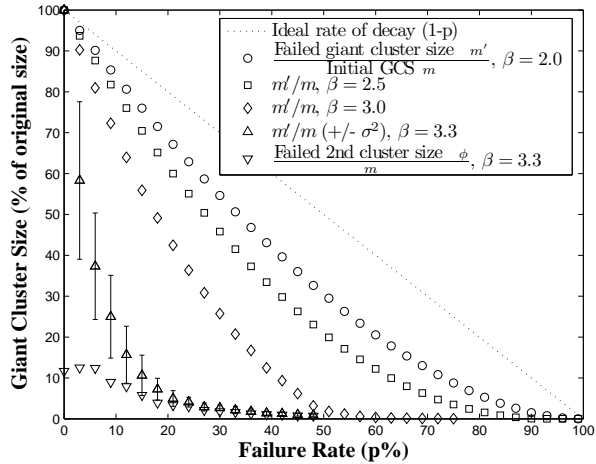}}}
\caption{The ratio of the size of the giant component remaining after
  failure to its size before failure, for $\beta = 2.0$, $2.5$, $3.0$,
  and $3.3$.  For $\beta = 3.3$ the standard deviation is shown.  For
  $\beta = 3.3$ the giant component is a fairly small fraction of the
  graph, and the giant component's disintegration is shown by
  observing the fractional size of the second largest component falls
  within this range by the predicted critical failure rate of
  $25.4\%$.  For $\beta = 3.0$ and $3.3$ the plots are truncated for
  clarity at $75\%$ and $50\%$, respectively.}
\label{failureAnalysis-gcs}
\end{figure}

We conclude with a graph emphasizing the decay of the giant component
itself, Figure~\ref{failureAnalysis-gcs}.  While $\beta' < \beta_0$
some constant-order (although potentially small) fraction of the
edge-bearing nodes in $G'$ will almost surely remain in a giant
component.  Let $m$ be the size of the giant component in $G$ and $m'$
its size in $G'$.  Then ideally $m' = (1-p)m$, but this is clearly not
the case.  This graph also magnifies the disintegration of the giant
component shown in Figure~\ref{sigSizeVsExp}, particularly in the
extreme case of $\beta = 3.3$.  In this case, the giant component is a
small (but constant-order) fraction of the graph to begin with, and as
the graph decays it is subject to greater uncertainty in its fate
(this curve was the only one with a substantive standard deviation),
so that it is not clear that it decays.  For this case, we include the
average size of the second largest component $\phi$, divided by $m$
(this value is graphically indistinguishable from zero in the other
three cases).  Through comparison of $m'/m$ with $\phi/m$ it appears
that the giant component has lost its status by the predicted critical
point.

\section{Conclusions and Future Work}\label{section-conclusions}

It is probably clear to any network administrator that static graph
theory has little to say about the resilience of the Internet to
random failures over time: the Internet is a massive computer network
with people responsible for repairing faults as they occur.
Nevertheless, the random simultaneous failure model is appropriate
when reasoning about the connectivity of randomly distributed
subpopulations in the network.  We believe we have shed some light on
the subtleties of a result that has been applied too generally.  By
explicitly considering orphans in the failure process of a PLN, and
considering graphs of minimum degree $1$, we have refined the Internet
resilience result of \citep{cohen00resilience} for finite networks.
Thus we have better estimated when such a network would completely
disintegrate as a function of its initial slope parameter and failure
rate.  In particular we have what we believe to be a more accurate
critical failure rate for the Internet, and we show that this is not
as resilient as originally suggested.

Because we are citing it so extensively, it is worth stating two
additional observations in \citep{cohen00resilience}, to avoid
confusion.  Cohen et al.\ make casual reference to the breakdown of
the SFN giant component under moderate failure levels when $\beta =
3.5$.  Although reasonable in general, in pure power-law networks
there is no giant component when $\beta > \beta_0 = 3.47875$
\citep{aiello00random}.  Also confusing may be their graph of the
fractional size of the giant component remaining as a function of the
failure rate $p$, when $\beta = 2.5$ (a graph analogous to our
Figure~\ref{failureAnalysis-gcs}).  This ratio is graphed in such a
way that it appears to very closely follow, and even exceed,
$(1-p)n$---in other words, the surviving component's size appears to
exceed the expected number of survivors.  In fact the figure graphs
this quantity {\em divided by $(1-p)$\/},\footnote{This discrepancy
  between figure and text has been confirmed by the authors in
  \citep{cohen05email}.}  and thus in fact matches our results in
Figure~\ref{failureAnalysis-gcs} for the values of $\beta$ they
present ($2.5$ and $3.5$).  This graph has been reproduced in several
places without elaboration \cite{albert02statistical}.

We have analytically considered these matters for finite PLNs under
the full range $0 < \beta < \beta_0$, but have not yet confirmed our
results in simulation for $\beta < 2$.  Observing how our results vary
for networks over several orders of magnitude larger and smaller also
remains to be done.  Doing so will require a more substantial
simulation than we have implemented.  Beyond pure power-law degree
distributions, a more general class of PLNs with rolloff and offset
terms should also be studied.  In particular, most real world networks
that approximate a power law exhibit a rolloff.  Finally, for $2 <
\beta < \beta_0$ we have been unable to find a derivation of the
fractional size of the giant component of the graph in the limit, and
such a formula would be of interest.  Extending the model to consider
assortative networks, conditional failure models, and other variations
that affect the critical failure threshold could lead to a number of
interesting new results.

\section{Acknowledgements}

We would like to thank Dr. Reuven Cohen for his review and comments of
the manuscript, and would also like to thank both Dr. Cohen and Dr. Lu
for their responses to our questions.

\bibliographystyle{elsart-num}
\bibliography{../bib/machine-learning,../bib/user-modeling,%
../bib/consensus-protocols,%
../bib/terran-pubs,../bib/hamilton-pubs,../bib/agents,../bib/mlrg}

\begin{thebibliography}{10}
\expandafter\ifx\csname url\endcsname\relax
  \def\url#1{\texttt{#1}}\fi
\expandafter\ifx\csname urlprefix\endcsname\relax\def\urlprefix{URL }\fi

\bibitem{cohen00resilience}
R.~Cohen, K.~Erez, D.~{ben Avraham}, S.~Havlin, Resilience of the {I}nternet to
  random breakdowns, Physical Review Letters 85~(21).

\bibitem{aiello00random}
W.~Aiello, F.~Chung, L.~Lu, A random graph model for massive graphs, in: STOC,
  Portland, OR, 2000, pp. 171--180.

\bibitem{albert02statistical}
R.~Albert, A.-L. Barab\'{a}si, Statistical mechanics of complex networks, Rev.
  Mod. Phys. 74 (2002) 47--98.

\bibitem{faloutsos99power-law}
M.~Faloutsos, P.~Faloutsos, C.~Faloutsos, On power-law relationships of the
  internet topology, in: SIGCOMM '99: Proceedings of the conference on
  Applications, technologies, architectures, and protocols for computer
  communication, ACM Press, New York, NY, USA, 1999, pp. 251--262.

\bibitem{chang04towards}
H.~Chang, R.~Govindan, S.~Jamin, S.~Shenker, W.~Willinger, Towards capturing
  representative {AS}-level {I}nternet topologies, Computer Networks Journal
  44~(6) (2004) 737--755.

\bibitem{csanyi04structure}
G.~Cs\'{a}nyi, B.~Szendr\H{o}i, Structure of a large social network, Physical
  Review 69~(036131).

\bibitem{reed04brief}
W.~J. Reed, A brief introduction to scale-free networks, unpublished work found
  at {http://www.math.uvic.ca/faculty/reed/draft\_1.pdf} at the Department of
  Mathematics and Statistics, University of Victoria (May 18 2004).

\bibitem{saroiu02measurement}
S.~Saroiu, P.~K. Gummadi, S.~D. Gribble, A measurement study of peer-to-peer
  file sharing systems, in: Proceedings of the Multimedia Computing and
  Networking (MMCN), San Jose, CA, 2002, pp. 156--170.

\bibitem{link05impact}
H.~Link, J.~Saia, T.~Lane, R.~Laviolette, The impact of social networks on
  multi-agent recommender systems, in: Proceedings of the Workshop on
  Cooperative Multi-Agent Learning (PKDD '05), Porto, Portugal, 2005.

\bibitem{newman03mixing}
M.~E.~J. Newman, Mixing patterns in networks, Physical Review E 67~(026126).

\bibitem{molloy95critical}
M.~Molloy, B.~Reed, A critical point for random graphs with a given degree
  sequence, Random Structures and Algorithms 6 (1995) 161--180.

\bibitem{molloy98size}
M.~Molloy, B.~Reed, The size of the giant component of a random graph with a
  given degree sequence, Combinatorics, Probability and Computing 7 (1998)
  295--305.

\bibitem{otter48number}
R.~Otter, The number of trees, Annals of Mathematics 49~(3) (1948) 583--599.

\bibitem{otter49multiplicative}
R.~Otter, The multiplicative process, Annals of Mathematical Statistics 20~(2)
  (1949) 206--224.

\bibitem{mohar91laplacian}
B.~Mohar, The laplacian spectrum of graphs, in: Y.~Alavi, G.~Chartrand,
  O.~Oellermann, A.~Schwenk (Eds.), Graph Theory, Combinatorics and
  Applications, Vol.~2, Wiley, New York, 1991, pp. 871--898.

\bibitem{achlioptas05bias}
D.~Achlioptas, A.~Clauset, D.~Kempe, C.~Moore, On the bias of traceroute
  sampling, or: Why almost every network looks like it has a power law, in:
  Proceedings of the 2005 Symposium on the Theory of Computation (STOC),
  Baltimore, MD, 2005, pp. 694--703.

\bibitem{awerbuch05improved}
B.~Awerbuch, B.~Patt-Shamir, D.~Peleg, M.~Tuttle, Improved recommendation
  systems, in: Proceedings of the Sixteenth Annual ACM-SIAM Symposium on
  Discrete Algorithms (SODA '05), SIAM, ACM, Vancouver, BC, CA, 2005.

\bibitem{kempe03maximizing}
D.~Kempe, J.~Kleinberg, {\'{E}}.~Tardos, Maximizing the spread of influence in
  a social network, in: Proceedings of KDD 2003, Washington, DC, 2003.

\bibitem{watts98collective}
D.~J. Watts, S.~H. Strogatz, Collective dynamics of 'small-world' networks,
  Nature 393 (1998) 440--442.

\bibitem{dorogovtsev04shortest}
S.~N. Dorogovtsev, J.~F.~F. Mendes, The shortest path to complex networks,
  arXiv:cond-mat/0404593 v1 (April 24 2004).

\bibitem{aiello01random}
W.~Aiello, F.~Chung, L.~Lu, A random graph model for power law graphs,
  Experimental Mathematics 10~(1) (2001) 53--66.

\bibitem{cohen01breakdown}
R.~Cohen, K.~Erez, D.~{ben Avraham}, S.~Havlin, Breakdown of the {I}nternet
  under intentional attack, Physical Review Letters 86~(16).

\bibitem{dorogovtsev01comment}
S.~N. Dorogovtsev, J.~F.~F. Mendes, Comment on `{Resilience of the Internet to
  Random Breakdowns}', Physical Review Letters 87.

\bibitem{bollobas85random}
B.~Bollob\'{a}s, Random Graphs, Academic Press, London, UK, 1985.

\bibitem{aiello05email}
W.~Aiello, L.~Lu, email with authors, confirming typographical errors in `{A
  random graph model for power law graphs,}' in Experimental Mathematics (May
  2005).

\bibitem{cohen05email}
R.~Cohen, email with author, confirming typographical errors in `{Resilience of
  the Internet to random breakdowns,}' in Phys. Rev. Lett. (September 2005).

\end{thebibliography}
\end{document}